\documentstyle{article}
\def\bea{\begin{eqnarray}}
\def\eea{\end{eqnarray}}
\def\nn{\nonumber}
\def\beq{\begin{equation}}
\def\eeq{\end{equation}}
\def\ba{\beq\new\begin{array}{c}}
\def\ea{\end{array}\eeq}
\def\be{\ba}
\def\ee{\ea}

\makeatletter
\newdimen\normalarrayskip 
\newdimen\minarrayskip 
\normalarrayskip\baselineskip
\minarrayskip\jot
\newif\ifold \oldtrue \def\new{\oldfalse}
\def\arraymode{\ifold\relax\else\displaystyle\fi} 
\def\eqnumphantom{\phantom{(\theequation)}} 
\def\@arrayskip{\ifold\baselineskip\z@\lineskip\z@
\else
\baselineskip\minarrayskip\lineskip2\minarrayskip\fi}
\def\@arrayclassz{\ifcase \@lastchclass \@acolampacol \or
\@ampacol \or \or \or \@addamp \or
\@acolampacol \or \@firstampfalse \@acol \fi
\edef\@preamble{\@preamble
\ifcase \@chnum
\hfil$\relax\arraymode\@sharp$\hfil
\or $\relax\arraymode\@sharp$\hfil
\or \hfil$\relax\arraymode\@sharp$\fi}}
\def\@array[#1]#2{\setbox\@arstrutbox=\hbox{\vrule
height\arraystretch \ht\strutbox
depth\arraystretch \dp\strutbox
width\z@}\@mkpream{#2}\edef\@preamble{\halign
\noexpand\@halignto
\bgroup \tabskip\z@ \@arstrut \@preamble \tabskip\z@ \cr}%
\let\@startpbox\@@startpbox \let\@endpbox\@@endpbox
\if #1t\vtop \else \if#1b\vbox \else \vcenter \fi\fi
\bgroup \let\par\relax
\let\@sharp##\let\protect\relax
\@arrayskip\@preamble}
\def\eqnarray{\stepcounter{equation}%
\let\@currentlabel=\theequation
\global\@eqnswtrue
\global\@eqcnt\z@
\tabskip\@centering
\let\\=\@eqncr
$$%
\halign to \displaywidth\bgroup
\eqnumphantom\@eqnsel\hskip\@centering
$\displaystyle \tabskip\z@ {##}$%
\global\@eqcnt\@ne \hskip 2\arraycolsep
$\displaystyle\arraymode{##}$\hfil
\global\@eqcnt\tw@ \hskip 2\arraycolsep
$\displaystyle\tabskip\z@{##}$\hfil
\tabskip\@centering
&{##}\tabskip\z@\cr}
\begingroup\ifx\undefined\newsymbol \else\def\input#1 {\endgroup}\fi

\def\Mp{M_{2p}}
\def\Wp{W_{2p+1}}
\def\di{d^{-1}}

\begin{document}

\setcounter{footnote}{1}
\def\thefootnote{\fnsymbol{footnote}}
\begin{center}
\hfill FIAN/TD-10/00\\
\hfill ITEP/TH-24/00\\
\hfill hep-th/0005244\\
\vspace{0.3in}
{\Large\bf On B-independence of RR Charges
}
\end{center}
\centerline{{\large A.Alekseev}\footnote{
Institute for Theoretical Physics, Uppsala University,
Box 803 S-751 08 Uppsala Sweden}, {\large
A.Mironov}\footnote{Theory Dept., Lebedev Physical Inst. and ITEP, Moscow,
Russia} and {\large A.Morozov}\footnote{ITEP, Moscow, Russia}}

\bigskip

\abstract{\footnotesize  Generalization of the recent Taylor-Polchinski argument
is presented, which helps to explain
quantization of RR charges in IIA-like theories
in the presence of cohomologically trivial $H$-fields.}

\begin{center}
\rule{5cm}{1pt}
\end{center}

\bigskip
\setcounter{footnote}{0}
\section{Introduction}

D-branes are sources of RR fields \cite{Dbr}.
In IIA theory a D$2p$-brane is coupled to the
RR gauge fields
$C^{(1)},  C^{(3)}, \dots,  C^{(2p+1)}$
(they are 1-, 3-, $\ldots$ and $(2p+1)$-forms respectively).
Their couplings to the brane depend on the
gauge invariant field 
$B_D = B + F$, where $F=dA$ is the tension
of $U(1)$ gauge field living on the brane, and $B$
is the gauge 2-form, pertinent for string theory.
The problem is that $B$ is not a tension and therefore
the integrals $\oint B_D^{k}$ over cycles in
brane's world sheet are not necessarily integer-valued
(only $\oint F^{k}$ are) and can not serve as charges \cite{BDS}.

In order to resolve this problem W.Taylor \cite{Taylor} and J.Polchinski
\cite{Pol} suggested 
to
substitute the naive formula
\cite{bract}

\be\label{1}
Q^{(1)} = \int_{M_2} B_D
\ee
for the $C^{(1)}$-charge of a topologically trivial closed D2-brane by:

\be
 Q^{(1)}  = \int_{M_2=\partial V_3} B_D - \int_{V_3} H
= \int_{M_2} F
\label{D2ch}
\ee
where $M_2 = \partial V_3$ is the position of D2-brane
and $V_3$ is any 3-volume with the boundary at $M_2$.
If the tension $H = dB$ is an exact 3-form, $dH=0$
(i.e. when $B$ is well defined), this expression
does not depend on the choice of $V_3$.
The argument of \cite{Taylor} in favor of (\ref{D2ch}) is essentially that
the bulk action mixes RR fields $C^{(2k+1)}$ with different $k$
and one needs to diagonalize the action before defining the
physical charges (as measured by remote probes).
The bulk action is diagonal in terms of the tensions
$G^{(2)}=d C^{(1)}$ and $G^{(4)}=d C^{(3)} - C^{(1)} \wedge H$
and the transformation from $C$'s to $G$'s of different
degree is given by a triangular matrix. So, the diagonalization
does not change the coupling to the highest RR field.
In order to define effective couplings to lower RR fields,
we should first integrate out all higher RR fields $ C^{(2m+1)} $
with $m>k$  and then read the coupling of $ C^{(2k+1)} $ to the brane.
Such diagonalization procedure gives rise to non-local contributions,
so that the resulting source terms are no longer concentrated on the
brane world sheet. However, the non-locality disappears for constant RR fields
which are used to probe the {\it charges}.

The purpose of the paper is to rephrase the reasoning of \cite{Taylor} and to
generalize it to arbitrary RR fields.
In the standard $d=10$ IIA string theory
the RR fields $C^{(2p+1)}$ and
$C^{(7-2p)}$ are dual to each other.
In this paper we ignore this restriction and
consider instead a different model which preserves the RR gauge symmetries but
all RR fields are independent of each other and the space-time
dimension is not specified. In this case, we show that  the effective coupling
of the RR fields to D-branes (after integrating out
higher RR fields) depends only on the tension $F$
of the U(1) gauge field on the brane. Since our interest is in the
$B$-dependence of the RR charges, we ignore the curvature-dependent corrections
\cite{GHM} and do not use the related $K$-theory formalism \cite{Kth1,Kth2}.

The paper is organized as follows.
We begin in sect.\ref{D2} with the case of
the D2-brane with $ C^{(3)}$ and $ C^{(1)}$ fields, considered
in \cite{Taylor}. Then, after a brief discussion of
the $d^{-1}$-operation in sect.\ref{dinv}, we proceed in
sect.\ref{gen} to generic consideration of $C^{(2p+1)}$ fields
(gauge odd forms) in the presence of $B^{(2k)}$ fields
(gauge even forms).
For cohomologically non-trivial fields $H$ and/or topologically non-trivial
branes, the gauge invariance produces constraints on the brane dynamics. They
are discussed in sect.5. Sect.6 addresses the ambiguities in RR charges arising
in these circumstances and their potential implications for quantum theory of
branes.

\section{The case of D2-brane \label{D2}}

The action of (the massless sector of) the IIA theory
can be obtained by dimensional reduction from the $d=11$
supergravity.  In $11d$ the bosonic sector consists of
the metric $G_{MN}$ and the  3-form ${\cal A}_{MNP}$,
subject to gauge transformations ${\cal A} \sim {\cal A} + d\sigma$ with any
2-form $\sigma$. The Lagrangian in $11d$ is:

\be
L_{11}=\sqrt{G}R(G) + |d{\cal A}|^2 +
{\cal A}\wedge d{\cal A}\wedge d{\cal A}
\ee
Here $|d{\cal A}|^2 = \sqrt{G} G^{M\tilde M}G^{N\tilde N}
G^{P\tilde P}G^{Q\tilde Q}
\partial_M{\cal A}_{NPQ}
\partial_{\tilde M}{\cal A}_{\tilde N\tilde P\tilde Q}$,
and the Chern-Simons term is independent of the metric.

After dimensional reduction to $10d$,
the $11d$ bosonic fields turn into:

\be
g_{\mu\nu} = G_{\mu\nu} +  C^{(1)}  _\mu  C^{(1)}  _\nu,
\  \ C^{(1)}  _\mu = G_{\mu,11}
\ee
and

\be
 C^{(3)}    _{\mu\nu\lambda} = {\cal A}_{\mu\nu\lambda},\  
B_{\mu\nu} = {\cal A}_{\mu\nu,11}
\ee
The $G_{11,11}$ component of the metric turns
into the exponential of the dilaton field, which is irrelevant
for our purposes and is omitted from all the formulas
in the present paper.
The gauge invariances (in addition to general coordinate
transformations in $10d$) are inherited from the 
general coordinate boosts in the 11-th dimension ($\epsilon^{(0)}$) and
from the gauge invariance of $A$ ($\Lambda$ and $\epsilon^{(2)}$),

\be
B \sim B + d\Lambda \ \ (\Lambda_\mu = \Lambda_{\mu,11}), \nn \\
 C^{(1)} \sim   C^{(1)} + d\epsilon^{(0)}, \nn \\
 C^{(3)} \sim   C^{(3)} + d\epsilon^{(2)} - d\epsilon^{(0)}\wedge B
\ee
The Lagrangian in $10d$ is of the form,

\be
L_{10}=\sqrt{g}R(g) + |dC^{(1)}|^2 +
\left|dC^{(3)} - C^{(1)}\wedge H\right|^2 +
B\wedge dC^{(3)}\wedge dC^{(3)}
= \nn \\ =
\sqrt{g}R(g) + | G^{(2)} |^2 + | G^{(4)} |^2  +
B\wedge  G^{(4)}  \wedge  G^{(4)}  - \nn \\ -
B\wedge B\wedge  G^{(4)}  \wedge  G^{(2)} 
- \frac{1}{3} B\wedge B\wedge B\wedge  G^{(2)}  \wedge  G^{(2)}  
+ \hbox{\ total\ derivative}
\ee
Here $H = dB$, $\ G^{(2p+2)} = d C^{(2p+1)} - C^{(2p-1)} \wedge H$,
and the total derivative appears from the transformation of the
Wess-Zumino term. 
The bulk Lagrangian $L_{10}$ contains terms mixing the fields
$C^{(1)}$ and $C^{(3)}$.
It can be diagonalized by the transformation
$C \rightarrow G$. However,
the inverse transformation $G \rightarrow C$ is non-local.

Introduce now even-dimensional D$2p$-branes located
at $\Mp$ with the world sheets $\Wp$. They contribute
the source (Wess-Zumino) terms to the action \cite{bract},

\be
\int_{W_1}  C^{(1)}   +\int_{W_3} ( C^{(3)} + C^{(1)}\wedge B_D)
+ \ldots
\label{stD2}
\ee
Here $B_D = B + F$ is a gauge invariant field,
which can be considered as a linear combination
of the gauge field $B$, living in the bulk, and
the tension of the $U(1)$ gauge field $F = dA$,
on the brane. Under the $\Lambda$-transformation,
the fields $A$ and $B$ transform as $A \sim A - \Lambda,
B \sim B + d\Lambda$. The tension $F$ is a closed
form (satisfies the Bianchi identity), $dF = 0$, so that
on the brane $dB_D = dB = H$.

The source (surface) terms (\ref{stD2}) are gauge
invariant up to total derivatives (i.e. for closed world-sheets 
$\partial \Wp = 0$).  However, they are not expressed
in terms of the $ G^{(2p+2)} $-fields. If one still wants to get
such an expression, a non-local formula occurs, no
longer concentrated on the world-sheets. 
For the time being we loosely use the operation $d^{-1}$,
its more adequate substitute will be discussed in the
following section \ref{dinv}.

So, $ C^{(1)}    = \di G^{(2)} $ and 

\be
 C^{(3)}     +  C^{(1)}   \wedge B_D =
\di G^{(4)}  +  C^{(1)}   \wedge B_D + \di( C^{(1)}   \wedge H)
\label{stra}
\ee
In order to obtain the $ C^{(1)}   $ (i.e. $ G^{(2)} $) {\it charge},
one needs to consider the coupling to D-brane of
(the time-component of) the {\it constant} $ C^{(1)}   $ field (emitted or
felt by a remote $C^{(1)}$-probe like a D0-brane). For constant $ C^{(1)}  $,
however, $\di( C^{(1)} \wedge H) = - C^{(1)}\wedge B$ and we obtain for $
Q^{(1)} $ exactly the formula (\ref{D2ch}). This is the argument of
\cite{Taylor}. It deserves mentioning that the Chern-Simons term in the
bulk Lagrangian vanishes for constant $ C^{(1)}   $.

The same argument is easily applied to the derivation
of $Q^{(2p-1)}$, the charge distribution
of $  C^{(2p-1)} $ (provided the bulk 
Lagrangian for $ C^{(2p+1)} $ fields is 
given by $| G^{(2p+2)} |^2$, as implied
by duality transformations $ C^{(2p+1)}  \rightarrow C^{(7-2p)}$).
However, in order to iterate our procedure and reach
lower $Q^{(2k+1)}$ with $k<p-1$ one can not keep
$C^{(2p-1)}$ constant. This motivates a more
detailed discussion of the operation $\di$ in the next section.

\section{Comment on the $d^{-1}$ operation \label{dinv}}

Because of the nilpotency property $d^2 = 0$ of exterior
derivative the $\di$ operator is not well-defined.
Its proper substitute is
the $K$-operation (de Rham homotopy), satisfying

\be
Kd + dK = 1
\ee
As required for $\di$, $K$ maps $k$-forms into
$(k-1)$-forms.
It is defined modulo $d$, for exact forms
$K(dV) = V - d(KV) = V \hbox{mod}(d)$. In application to integrals over
topologically trivial surfaces, $K$ can be given, for example,
by the following explicit construction (known for physicists from the
fixed-point gauge formalism of the early days of gauge theories, see, e.g.,
\cite{UFN}). For a 1-form $A(x)$, the 0-form

\be
KA(x) = \int_0^1 A_\mu(tx)x^\mu dt
\ee
and in general for a $k$-form $A(x)$

\be
\left(KA(x)\right)_{\mu_1\mu_2\ldots\mu_{k-1}} =
\int_0^1 A_{\mu_1\ldots\mu_k}(tx)x^{\mu_k} t^{k-1}dt
\ee
or

\be
\int_{S_{k-1}} KA = \int_{C(S)_{k}} A
\ee
where $C(S)$ is a {\it cone} with the base $S$
and the vertex somewhere outside $S$. Of course, the $K$ operation
depends on the choice of this vertex, but
this dependence results in a gauge transformation
for gauge forms.
The $n$-fold application of $K$  builds up
an $n$-dimensional simplex $C^n(S)$ over the surface $S$.
For $k_i$-forms $A_i$, $\sum k_i=k$  one obtains,

\be
\int_{S_{k-n}} K
\left(A_{k_1} K\left(A_{k_2} K\left(\ldots
KA_{k_n}\right)\right)\right) =
\int_{C^n(S)_k} A_{k_1}\wedge A_{k_2}\wedge \ldots \wedge A_{k_n}
\ee
Coming back to the surface term (\ref{stra}), one can
note that $\di$ in this expression can be safely substituted
by $K$, because $Kd C^{(3)}\sim C_3$ modulo gauge transformation.
At the same time, in variance with (\ref{stra}),

\be
\int_{W_3}\left( C^{(3)} + C^{(1)} \wedge B_D\right) =
\int_{W_3} \left(K G^{(4)} +  C^{(1)}   
\wedge B_D + K( C^{(1)}   \wedge H)\right)
= \nn \\ =
\int_{W_3}  C^{(1)}   \wedge B_D +
\int_{C(W_3)} \left( G^{(4)} + C^{(1)} \wedge H\right) 
\label{stramod}
\ee
is a well defined expression for non-constant fields $C^{(1)}$.

\section{Generic RR fields \label{gen}}
At this point we deviate from conventional IIA theory,
and assume that the RR field is an odd gauge poli-form in the bulk,

\be
C = \sum_{k=0} C^{(2k+1)}
\ee
with no interrelations on $C^{(2k+1)}$ imposed.
The background is given by
an even gauge poli-form $B$ in the bulk,

\be
B = \sum_{k=1} B^{(2k)}
\label{Bf}
\ee
and by an odd gauge poli-form $A$ on the D-brane world-sheets,

\be
A = \sum_{k=1} A^{(2k-1)}, \ \ \
F = dA = \sum_{k=1} dA^{(2k-1)},\ \ \ B_D\equiv B+F
\ee
The gauge symmetries are as follows,

\be
C \sim C + e^{-B}d\epsilon, \ \  \
B \sim B + d\Lambda,  \ \ \
A \sim A - \Lambda + d\alpha,
\ee
where

\be
\epsilon = \sum_{k=0}\epsilon^{(2k)}, \ \ \
\Lambda = \sum_{k=1}\Lambda^{(2k-1)}
\ee
are even and odd poli-forms in the bulk respectively, and 

\be
\alpha = \sum_{k=0} \alpha^{(2k)}
\ee
is an even poli-form on the brane world sheet.
The tension $H = dB$ of the $B$-field and the tension $G = dC - CH$
of the RR field are gauge invariant bulk fields.
As before, the combination $B_D=B+F$ is a gauge invariant
field on the brane world-sheet.

In analogy with the IIA theory we assume the following
Lagrangian of RR fields
(in neglect of possible Chern-Simons terms) in the bulk,

\be
|G|^2 =
\left| dC - C dB\right|^2 =
\sum_i \left| dC^{(2i+1)} - C^{(2i-1)}\wedge dB\right|^2 .
\ee
Here and in what follows we often omit the sign of wedge product,
meaning that contractions with the help of the metric
involve either the squares $|\ |^2$ or the Hodge
operation $*$. In particular $e^{-B} = e^{-\wedge B}$,
$CH = C\wedge H$ etc.

The Wess-Zumino contributions of D-branes are \cite{bract}\footnote{
We neglect here the curvature terms, which are given by
a pullback of the factor $\sqrt{\hat A(R)}$ in the integrand
\cite{GHM,Kth1}, where the $\hat A$-genus is the one, appearing
in index theorems, like
$
\ {ind}(D) = \int \hat A(R)\ \hbox{tr}\ e^{-F}
$.}:

\be
\sum_p \oint_{\Wp} C  e^{B_D}  =
\sum_p\left(
\sum_{i=0} \frac{1}{i!} \,\oint_{\Wp} C^{2p+1 - 2i}\wedge B_D^{i}
\right) =
\sum_{k,i} \frac{1}{i!} \, \oint_{W_{2k+2i+1}} C^{(2k+1)}\wedge B_D^{i}  .
\label{surtr}
\ee

According to our general strategy we should rewrite the
source term (\ref{surtr}) in terms of  $G$, which diagonalizes
the bulk action. The resulting expression will essentially
contain the $K$ operation, but formulas for the RR {\it charges}
are obtained at constant RR fields
when the expression becomes local.
Formally, in order to express $C$ through $G$, one
needs to solve the equation

\be
G = dC - CH = (d + H)C=\left(e^{-B}de^{B}\right)
\ee
For our purposes, we need a formula with $K$-operation acting directly on $C$.
Since up to a gauge transformation $C \sim K(dC)$, we
write $C \sim KG - KHC$, and

\be
C \sim \frac{1}{1+KH} KG = KG - K(H\wedge KG) + K(H\wedge K(H\wedge KG))
+ \ldots
\ee
Since $K$ lowers the rank of the form by one and multiplication
by $H$ raises it by three or more (note that there is no terms
with $k=0$ in (\ref{Bf})), this formula contains only
finitely many terms for $C$ of a given finite rank.
Finally, the source term turns into:

\be
\sum_p \int_{\Wp} Ce^{B_D} =
\sum_p \int_{\Wp} e^{B_D}\frac{1}{1+KH}KG
\label{genfor}
\ee
Now the entire Lagrangian is diagonalized in terms of the fields $G$ and we
can proceed to the definition of effective couplings.

\subsection*{Examples.}

For a D2-brane we have

\be\label{27}
\int_{W_3} \left(C^{(3)} + B_D^{(2)}C^{(1)}\right) =
\int_{W_3} (1 + B_D + \ldots ) (1 - KH + \ldots)
(KG^{(2)} + KG^{(4)} + \ldots) = \nn \\ =
\int_{W_3} \left(KG^{(4)} + B_D KG^{(2)} - K(H (KG^{(2)}))\right) =
\int_{W_3}  KG^{(4)} + \int_{W_3}\left(B_D C^{(1)} +
K(C^{(1)}H)\right)
\ee
(we used the fact that $KG^{(2)} \sim C^{(1)}$). This reproduces
eq.(\ref{stra}).

Similarly, for a D4-brane:
\be
\int_{W_5} 
\left(1 + B_D^{(2)} + B_D^{(4)} + \frac{1}{2}(B_D^{(2)})^2 + 
\ldots\right)
\cdot \nn \\ \cdot 
\left(1 - K H^{(3)} - K H^{(5)} + K H^{(3)} K H^{(3)} + \ldots\right)
\left( KG^{(2)} + KG^{(4)} + KG^{(6)} + \ldots \right) = \nn \\ =
\int_{W_5} KG^{(6)} +
\int_{W_5} \left(B_D^{(2)}KG^{(4)} - K(H^{(3)}(KG^{(4)}))\right)
+ \nn \\ +
\int_{W_5} \left(\left(B_D^{(4)} + \frac{1}{2}(B_D^{(2)})^2\right)
 KG^{(2)} - K(H^{(5)} (KG^{(2)})) +
K\left(H^{(3)} K(H^{(3)} KG^{(2)})\right)\right)
\label{4br}
\ee

In order to read off the coupling of $C^{(5)}$
from (\ref{4br}) one notes that this field
enters only through $G^{(6)}$ and 
the $K$-operation can be explicitly applied
to this term: $\int_{W_5} K(dC^{(5)}) = \int_{W_5} C^{(5)}$.

The procedure to define the $C^{(3)}$-``charge'' literally repeats the one for
the $C^{(1)}$-charge of D2-brane. First, we assume (following \cite{Taylor})
that $G^{(6)}$ is already integrated out, thus one does {\it not}
look at $C^{(3)}$ in $G^{(6)} = dC^{(5)} - C^{(3)}\wedge H^{(3)}
- C^{(1)}\wedge H^{(5)}$. Then $C^{(3)}$ enters (\ref{4br})
only through $G^{(4)}$. Second, one should put $C^{(3)} = const$,
then the $K$ operation can be applied explicitly and give:

\be
\int_{W_5}\left(B_D^{(2)}C^{(3)} - K(H^{(3)} C^{(3)})\right)
\stackrel{C^{(3)}=const}{=} \
\int_{W_5} (B_D - B) C^{(3)} = \int_{W_5} C^{(3)} F
\ee

Applying the same procedure to the case of the $C^{(1)}$-charge, i.e.
omitting the terms with
$G^{(6)}$ and $G^{(4)}$, putting $KG^{(2)}=C^{(1)} = const$
and applying $K$ explicitly) we obtain from (\ref{4br}):

\be
KG^{(2)}=\int_{W_5}\left( \left(B_D^{(4)} + \frac{1}{2}(B_D^{(2)})^2\right)
 C^{(1)} - K(H^{(5)} C^{(1)}) +
K\left(H^{(3)} K(H^{(3)} C^{(1)})\right)\right)
= \nn \\ \stackrel{C^{(1)}=const}{=} \
\int_{W_5}\left( \frac{1}{2}(B_D^{(2)} - B^{(2)})^2 +
(B_D^{(4)} - B^{(4)})\right) C^{(1)} =
\int_{W_5} \left(\frac{1}{2} (F^{(2)})^2 + F^{(4)}\right)
C^{(1)}
\ee
so that for a  D4-brane

\be
Q^{(1)} = \int_{M_4} \left(\frac{1}{2} (F^{(2)})^2 + F^{(4)}\right)
\ee

\subsection*{$B$-independence of RR charges.}

These examples illustrate the general prescription to define the
effective D-brane coupling to
the constant RR field $C^{(2k+1)}$:
in (\ref{genfor}) neglect all the contributions of
$G^{(2m+2)}$ with $m \neq k$ and in the term with $KG^{(2k+2)}$
put $C^{(2k+1)} = const$. Then, it appears that the non-localities remaining in
the $K$-operation can be explicitly eliminated, the contributions of $B$ to
$B_D$ are completely canceled, and the final answer depends
only on $F$. It deserves noting that while $F$ itself is
not gauge invariant, the integrals $\oint F^k$ over compact $2k$-cycles are.

The way to prove these claims is actually clear from
above examples. It is enough  to note that whenever
$KG$ in (\ref{genfor}) is substituted by a {\it constant} $C$ to obtain,

\be
\frac{1}{1+KH} C \ \stackrel{C= const}{=}\
\sum_{j=0} (-)^j (KH)^j C = \sum_{j=0} \frac{(-)^j}{j!}
B^jC = e^{-B}C .
\ee
Then, since $e^{B_D-B} = e^{F}$,
the source term (\ref{genfor}) becomes  $B$-independent,

\be
\sum_p \int_{\Wp} e^{F} C .
\ee
This formula for the effective coupling of a D-brane to the constant
RR gauge fields is the main result of the paper. It shows
that only the integral 2-form $F$ contributes to the properly defined RR
charges, in analogy to the claim of Taylor-Polchinski for D$2$-branes.

\section{Implications of gauge non-invariance of the
source terms \label{zmodes}}

Though both the bulk and the world-sheet actions are
expressed through the gauge-invariant $G$-fields, the non-locality
of the operation $K$ can -- and does -- (slightly)
diminish the gauge invariance
of the source (world-sheet) terms. 
This leads to additional constraints imposed on the
brane configurations contributing to the functional integral over RR fields.

For example,
the bulk action is invariant under

\be
C \rightarrow C - \varepsilon\  de^{- B}
\label{varep}
\ee
with $\varepsilon = const$.
(Indeed,
$G \rightarrow d\varepsilon\  de^{-B} - \varepsilon\ de^{-B} dB$,
and the first term vanishes for constant $\varepsilon$,
while the second one vanishes because $dBdB = dB \wedge dB = 0$.)
However, the
boundary (source) term is not invariant:

\be
\delta \oint C e^{B_D} =
\oint \varepsilon\ dB e^{B_D-B} =
\oint \varepsilon\ dB e^{F}
\ee
Integration over the zero-mode $\varepsilon$
of the full action  provides  the
constraints,

\be
\forall k\geq 0\ \ \ \oint_{V_{3+2k}} F^k dB = 0
\label{cons}
\ee
for any cycle in the D$2p$-brane's world-volume.
In particular, for a D2-brane one gets $\oint_{W_3} H = 0$,
i.e. the D2-brane can not wrap around a source of $H$-field
(like an NS brane) in its time-evolution (such trajectories
do not contribute to the functional integral).

Generalizing (\ref{varep}) to
$$
C \rightarrow C + \varepsilon e^{-B} B^m dB
$$
one obtains extra constraints:
$$
\oint F^kB^mdB = 0
$$
or their linear combinations $\oint F^kB_DdB_D$.
At the same time, the $B$-independent gauge transformation
$$
C \rightarrow C + \varepsilon
$$
with constant $\varepsilon$ can be easily excluded from
the gauge group from the very beginning, and therefore the
integrals $\oint F^k$ without $B$-fields
need not vanish, i.e. dynamics of RR fields eliminates
wrappings around ``magnetic'' sources of $B$ fields, but not
of $F$ fields.

\section{RR charge in cohomologically non-trivial $H$}

When $H$ is cohomologically nontrivial, expressions
like (\ref{D2ch}), (\ref{stra}) and (\ref{27}), in variance with the
naive (\ref{1}) are ill-defined. Still, it seems that
they are adequate for description of the actual
situation, and the ambiguities may appear to have physical
significance.
An ambiguity  problem arises if there are non-contractable
3-cycles in the space-time with non-vanishing $\oint H$
(e.g., in Calabi-Yau compactifications) or sources of
``magnetic'' $H$-fields, where $dH \neq 0$
(e.g., the NS5-branes). Similar problems occur for
topologically-nontrivial D-branes, i.e. when
$M_2 \neq \partial V_3$: then (\ref{stra}) and (\ref{27}), in variance
with (\ref{D2ch}) are still applicable, but ambiguous.

In order to understand what happens in such situations,
one can analyze a 1-dimensional toy-example. For a puzzle,
involving realistic D2-branes, see ref.\cite{parad}.
Consider the 1-dimensional Gaussian theory on a circle $S_1$ whose
partition function is

\be
\int {\cal D}C(x) \ \exp \left(\frac{i}{2\pi}
\oint_{S_1} \left|\frac{\partial C}{\partial x} -
\sigma\frac{\partial b}{\partial x}\right|^2 \ + \
C(x_1) - C(x_2) \right)
\label{1din}
\ee
Here $C(x)$ models the $C^{(3)}$ field in the bulk,
$b(x)$ and $h = \partial b/\partial x$ -- the $B$ and $H$
fields respectively, $\sigma$ plays the role of the
constant $C^{(1)}$ in the bulk, and the source term
$\int C^{(3)}$ (dipole charge)
on the brane world-sheet is imitated by $C(x_1) - C(x_2)$.
Cohomologically non-trivial $h$, such that $\oint_{S_1}
h(x)dx \neq 0$, arises when $b(x)$ is non-periodic,
$\Delta b = b(x+1) - b(x) \neq 0$, with periodic $h(x)$.

The answer for the Gaussian integral (\ref{1din}) (with
eliminated gauge zero-mode $C(x) = const$) is

\be
\exp \left[2\pi i\sigma \left(b(x_1) - b(x_2)\right)\right]
\ee
and for non-periodic $b(x)$ it is an ambiguous function of
$x_1$ (with $x_2$ fixed). If there were no $\sigma$
(no $C^{(1)}$ field) coupled to $h(x)$ in the bulk,
one could have handled this problem by imposing a Dirac quantization
condition on $b$: $\Delta b = integer$. However, in our model $\sigma$
is a dynamical variable and, even if constant,
can vary continuously. In other words, coupling to RR
fields makes the phase ambiguity physically relevant.
As usual in such situations, the physical state is not
defined solely by the current {\it position} of the brane
(by the points $x_1$ and $x_2$ on $S_1$), it also
depends on its pre-history: given original state at some moment, one can
obtain at another moment the physically distinct states with the same $x_1$ and
$x_2$, differing by the number of times the $x_1$ and $x_2$ wrapped around the
circle. The situation may be reminiscent of the theory of anyons. The state of
such a system is not fully determined by current positions of the anyons, it is
also labeled by an attached (invisible) braid. Moreover, the ambiguity is
not automatically resolved by simply summing over all the
pre-histories: dynamical constraints, discussed in the previous section lead to
a superselection rule, preventing dynamical interference of topologically
different pre-histories.

\section{Acknowledgements}

We are indebted to the ESI, Vienna, for the hospitality
and support which made this cooperation possible.
Our work is partly supported by grants:
NFR F 674/1999, INTAS 96-0196, INTAS 99-01705 (A.A.), RFBR 98-01-00328, INTAS
97-0103 (A.Mir), RFBR 98-02-16575 (A.Mor.), CRDF \#6531 (A.M.'s)
and the Russian President's grant 00-15-99296 (A.Mor.).

\end{document}